\documentclass{article}
\usepackage{graphicx}

\begin{document}
\noindent {\bf \large On Debye temperature anomaly observed in Ge-Se-Ag glasses}
\vskip1.5cm
\noindent {\bf Ashok Razdan}

\noindent {\bf Astrophysical Sciences Division }

\noindent {\bf Bhabha Atomic Research Centre }

\noindent {\bf Trombay, Mumbai- 400085 }
\vskip 1.5cm
\noindent {\bf \large Abstract:}

Anomalous values of Debye temperature have been obtained from
recoil free factor measurements Ge-Se-Ag glasses recently [1].
In the present  paper we show that this anomaly  may arise due to the presence of anharmonic potential at the
high spin ferrous site. We use q Lamb Mossbauer factor and anharmonic Lamb
Mossbauer factor to study this anharmonicity

\vskip 1.5cm
PACS Numbers : 63.70.+h, 63.20.+Ry,76.80.+y,33.25.+k
\vskip0.5cm
Keywords: Lamb Mossbauer factor, deformation, anharmonicity, Chalcogenide glasses, Debye temperature

\vskip 1.5cm
email:akrazdan@barc.gov.in
\vskip 0.5cm
country: India, Place Mumbai
\vskip 0.5cm
telephone:91-022-25591798
\vskip 0.5cm
Fax: +9122 25505151

\newpage
\noindent {\bf \large Motivation:}

Chalcogenide glasses have very interesting microscopic and macroscopic properties.
One such example of chalcogenide glasses Ge-Se-Ag systems, has been studied by Arcondo et.al.[1] recently.
From Mossbauer spectroscopy measurements of Ge-Se-Ag systems, it is clear that ferrous is present in low spin and
high spin states. Low spin $Fe^{2+}$ corresponds to octahedral co-ordination environment and
high spin $Fe^{2+}$ corresponds to distorted octahedral co-ordination environment.
It is also observed that two environments correspond to two Debye temperatures. The Debye
temperatures of  $\Theta_D$ = 290 K corresponds to high spin $Fe^{2+}$ state and   
$\Theta_D$ = 370 K corresponds to low  spin $Fe^{2+}$ sites of Ge-Se samples with $Ag_{10}$.
These two values of $\Theta_D$ indicate presence of two phases.
The XRD  of Ge-Se-Ag glasses shows single or homogeneous phase [1]. The presence of
single glass transition also indicates presence of single [1] or homogeneous phase.  

In this paper we show that Mossbauer recoil free factor data can also be explained using single value 
of Debye temperature  thereby agreeing with XRD and glass transition results.
It is important to note that Mossbauer recoil free factor is very sensitive to the
nature of potential. The standard form of Lamb Mossbauer factor (equation 8) has been derived 
for harmonic potential.
We presume  that at low spin site of $Fe^{2+}$ that there is presence of harmonic potential
and equation (8) can be applied to calculate Debye temperature. At high spin site of $Fe^{2+}$, 
we assume there is anharmonic potential  . The presence of
anharmonic potential at high spin site modifies the temperature dependence of
normal f-factor, thereby resulting in
biased calculation of Debye temperature when we use equation (8)
which holds only for harmonic potential. Again it is natural to assume that the presence
of anharmonic potential at high spin $Fe^{2+}$  is due to the distorted octahedral co-ordination environment.

In the following we use deformed Lamb Mossbauer factor and conventional anharmonic Mossbauer recoil 
free factor to study anomalous nature of $\Theta_D$ at $Fe^{2+}$  high spin site. 
 
\noindent{ \bf \large Deformed Lamb Mossbauer factor:}

Quantum deformation of algebras and groups have resulted in deformed analogue or commonly known as
q analogue of harmonic oscillator. Statistical properties of q Bose gas has been studied in detail
[2,3] and has found wide range applications in nuclear physics, molecular physics, non linear optics
and condensed matter physics [4] and references therein. It has been observed that eigen values
of q-harmonic oscillators are not equally spaced .  The spectrum of q harmonic oscillator has been
found to be similar to the spectrum of anharmonic oscillator [5,6]. Atroni et. al. [7] have interpreted the
physical significance of q parameter as a measure of degree of anharmonicity. It has been found
that parameter q in q specific heat [8] is a measure of degree of anharmonicity.  It has been argued that
specific heat measurements in superfluids can be explained only if phonons follow q deformation algebra.
So by studying q version of Lamb Mossbauer factor it has been possible to study anharmonicity in
q f-factor[9].

The Hamiltonian of q harmonic oscillator is defined by
\begin{equation}
H=\frac{1}{2} \hbar \omega (a^{+}a + aa^{+})
\end{equation}
where $a^{+}$ and a are creation and Annihilation q-Bose Operators which satisfy the following commutation 
relation
\begin{equation}
aa^{+} -q a^{+} a =1
\end{equation}
q is deformation index and
\begin{equation}
a^{+} a = [N]
\end{equation}
is number operator. The square bracket is a q-number. For any number r, the q number defined as
\begin{equation}
[r]= \frac {q^r - \frac{1}{q^r}}{ q -\frac{1}{q}}
\end{equation} 

Average occupation Number
\begin{equation} 
n_q=\frac {\sum_{0} ^{\infty} T exp(-T)}{ \sum_{0} ^{\infty} exp(-T)}
\end{equation}
where
\begin{equation}
T=\frac{1}{2}  ([n+1] +[n] -1)
\end{equation}
The Lamb Mossbauer Factor using equations generalized quantum mechanics
\begin{equation}
log f_0 = \frac{-E_{\gamma} ^{2}}{3mc^{2}} \int_{0}^{\omega_{max}}   (\frac{1}{2}+ n_q)
(\frac{\hbar}{k})^{3}   \frac{\omega_i^{2} d \omega_i}{\Theta_D ^{3} \hbar \omega_i}
\end{equation}
where x= $\frac{\hbar \omega}{k T}$ and $ \hbar \omega_{max}$= $ k \Theta_D$. The  equation(7)
holds for real and positive value of q.
It has been shown that for deformed Lamb Mossbauer factor parameter q decides the
nature of f-factor (in equation 7) dependence on temperature.  Deformed Lamb Mossbauer factor
has been used to study anharmonicity [9] in antimony, prussian blue analogues. It
has also been used to study anharmonicity in superconductors like $Nb_3 Sn$ [10] and YBaCuO [11].

In the limit of q $\rightarrow$ 1, equation (7) reduces to
\begin{equation}
log f_0 = \frac{-3E_{\gamma}^{2}}{mc^{2} \Theta_D} [ \frac{1}{4}+( \frac{T}{\Theta_D})^2
 \int_{0}^{ \frac{\Theta_D}{T}} \frac{xdx}{e^x-1}]
\end{equation}

This is well known expression for Lamb Mossbauer factor using normal quantum mechanics.

\noindent{\bf \large Conventional anharmonic approach:}

Marradudin and Flinn have derived a relationship which describes the effect of anharmonicity on
the recoil free factor f. This relationship is given as [12,13]
\begin{equation}
ln f = \frac{ -6 RT}{\Theta_D^2}  (1+\epsilon T+ ...)
\end{equation}

In above equation, R is the recoil energy of nucleus, $\epsilon$ is the anharmonic coefficient. 
For $^{57}Fe$ nucleus, R =22.6 K. The anharmonic coefficient $\epsilon$ can be theoretically calculated
by using Maradudin and Flinn theory. Experimentally $\epsilon$ can be measured from recoil free factor
versus temperature curves. 

The similarity between q Lamb Mossbauer factor and conventional anharmonic approach has
been very well studied in [11,10]. At low temperatures this similarity is not good but above
a 'particular temperature' this similarity improves sharply. The value of this 'particular temperature' is
strongly dependent on the value of the Debye temperature. For low Debye temperature, the similarity
begins at lower temperature [11].
 
\noindent{ \bf \large Discussion and results:}

The relationship between normal recoil free factor and Debye temperature is very interesting
at all temperatures [14]. For
high Debye temperature, f-factor falls slowly with the increase of temperature. 
For small  Debye temperature  f-factor
falls sharply even with small  increase of temperature.
Lamb Mossbauer factor reduces sharply even at 0 K [14] for a solid with low Debye temperature.
    
Earlier we have  shown that both deformed Lamb Mossbauer and anharmonic Lamb Mossbauer factor
affect  recoil free factor 's dependence on temperature very strongly. It is clear that both 
q and $\epsilon$ affect
f-factor in almost similar manner i.e. larger the value of q or $\epsilon$ , larger is their effect on f-factor and
vice versa. Again it is important to note that value of $\epsilon$ needed to affect the change in
f-factor depends on value of Debye temperature i.e. larger the value of Debye temperature, larger is the
value of $\epsilon$ needed to affect the nature of f-factor and vice versa. This is  also  true of
q parameter [10,11]. Thus anharmonic coefficient $\epsilon$  shares a similar relationship with Debye temperature as
deformed parameter q does. We have shown it earlier that there is a qualitative (if not quantitative)
similarity in nature of q and $\epsilon$. Both q and $\epsilon$ represent anharmoncity and anharmonicity affects
recoil free factor by decreasing it more rapidly than in the case of a normal Debye solid.

Figure 1 depicts four curves (a-d). The curve 'a' corresponds to Debye temperature of $\Theta_D$= 370 K and
curve 'b' corresponds to $\Theta_D$= 290 K. Both curves 'a' and 'b' have been obtained using normal Lamb Mossbauer factor
model of equation(8). Curves 'c' and 'd' has been obtained by using q-Lamb Mossbauer factor of equation (7). 
Curves 'c' and 'd' have been obtained for Debye temperature 
$\Theta_D$=370 K. For curve 'c' q=1.98 and for curve 'd' q=2.04 . It is clear that curves 'c' and 'd'
obtained for $\Theta_D$=370 K surround  curve 'b' which has been obtained for $\Theta_D$=290 K. 
Curve 'b' represents f-factor of high spin $Fe^{2+}$ site.
Curve 'b' is also very similar to curves 'c' and 'd' (for T $>$100K) which correspond to Debye temperature of $\Theta_D$=370.

Figure '2' depicts four curves (a-d).
The curve 'a corresponds to Debye temperature of $\Theta_D$= 370 K and
curve 'b' corresponds to $\Theta_D$= 290 K. Both curves 'a' and 'b' have been obtained using normal Lamb Mossbauer factor
model of equation(8). Curves 'c' and 'd' have
been obtained using conventional anharmoic approach of equation (9). 
Curves 'c' and 'd' have been obtained for Debye temperature
$\Theta_D$=370 K. For curve 'c' $\epsilon$ =0.0031 and for curve 'd' $\epsilon$=0.0045 . It is clear that curves 'c' and 'd'
obtained for $\Theta_D$=370 K surround  curve 'b' which has been obtained for $\Theta_D$=290 K.
Curve 'b' represents f-factor of high spin $ Fe^{2+}$ site.
Curve 'b' is also very similar to curves 'c' and 'd' (for T $>$100K) which correspond to Debye temperature of $\Theta_D$=370.

Curve 'a' in figures 1 and 2, corresponds to Debye temperature $\Theta_D$= 370K, 
representing f-factor of low spin $Fe^{2+}$ site.
At low spin site of $Fe^{2+}$ there is presence of harmonic potential for which q=1 or 
$\epsilon$=0.
Thus we can say Curve 'b' in figures 1 and 2,representing f-factor of high spin $ Fe^{2+}$ site,
also corresponds to Debye temperature $\Theta_D$= 370 K for which there is presence of anharmonic 
potential corresponding to  q= 2.02 $\pm$ 0.03 or  $\epsilon$= 0.0038 $\pm$ 0.0007
Thus curve 'b' is indeed a case of anharmonic behaviour because $\epsilon \ne 0$ or q $\ne$ 1 and
curve 'a' is a case of harmonic behaviour because $\epsilon$ =0,  or q =1. However, in the
same lattice one site cannot experience harmonic behaviour ( i.e. q=1, or $\epsilon$ = 0) and
another site cannot experience anharmonic behaviour ( i.e. q $\ne$ 1 or $\epsilon \ne 0$ ) for
similar thermal conditions. So Ge-Se-Ag glasses do not experience whole lattice anharmonicity.
The present problem is not a case of anharmonicity due to thermal expansion but a localized  effect
due to the presence of anharmonic potential well in the vicinity of high spin $ Fe^{2+}$ due to
the presence distorted octahedral co-ordination environment.
The distorted octahedral co-ordination environment may  create
anharmonic potential at high spin site, modifying  the f-factor,thereby resulting in
biased calculation of different Debye temperature using standard Lamb Mossbauer factor.
In the present studies the value of q = 2.02 or $\epsilon$ = 0.0038 may be quantifying
the anharmonicity due to the presence of this potential well.
Almost a similar scenario has been observed in high temperature properties of YBaCuo supercondutors [11].
 
Anharmonicity in Mossbauer spectroscopy is either a high temperature or low temperature phenomena. High
temperature anharmonicity in Lamb Mossbauer occurs when small to moderate deviations take place
in the parabolic nature of potential. These deviations exist in only high temperature region. 
In contrast to this low temperature anharmonicity arises due
to presence of non-parabolic potential  and will be appreciably present as anharmonic term even at
absolute zero temperature.
Recoil free factor is the product [15] of harmonic and anharmonic terms. The harmonic term is subject
to linear temperature dependence while the anharmonic terms exhibits little or no temperature dependence.
Therefore the effect of low temperature anharmonicity is to displace the harmonic curve by large values
at all temperatures. 
It may be interesting to study f-factor dependence of Ge-Se-Ag glasses in low temperature region to establish 
true nature of anharmonicity.

\noindent{ \bf \large Conclusions:}
In this paper we have shown that  Debye temperature anomaly observed in Ge-Se-Ag glasses can be
solved by assuming presence of the anharmonic potential at the high spin ferrous site. This anharmonic 
potential at ferrous site may arise due to the distorted octahedral co-ordination enviornment. 

\noindent{ \bf \large  References:}
\vskip 0.5cm
[1]  B.Arcondo, M.A.Urena, A.Piarristeguy, A.Pradel and M.Fontana,  Physica B 389(2007)77
\vskip 0.1cm
[2]  A.J.Macfarlane, J.Phys. A 22(1989)4581
\vskip 0.1cm 
[3]  L.C.Biedenharn, J.Phys. A 22(1989)L873
\vskip 0.1cm 
[4]  D.Galetti,B.M.Pimental,C.L.Lima,J.T.Lunardi, Physica A 242(1997)501
\vskip 0.1cm
[5]  J.L.Birman Phys. Lett. A 167(1992)362
\vskip 0.1cm
[6]  D.Banatsos and C.Darskaloyannis,
Phys. Lett. B 278(1992)1
\vskip 0.1cm
[7]  M.Atroni,J.Zang and J.L.Birman, Phys. Rev. A 47(1993)2555
\vskip 0.1cm
[8]  S.Zhang, Phys. Lett. A 202 (1995)18
\vskip 0.1cm
[9]  A.Razdan, Phys. Status Solidi B 203(1997)337
\vskip 0.1cm
[10] A.Razdan, Pramana, Journal of Physics 54(2000)871
\vskip 0.1cm
[11] A.Razdan, Hyperfine Interactions 122(1999)309
\vskip 0.1cm
[12] A.A.Maradudin and P.A.Flinn, Phys. Rev. 126(1962)9
\vskip 0.1cm
[13] A.A.Maradudin and P.A.Flinn, Phys. Rev. 129(1962)252
\vskip 0.1cm
[14] V.G.Bhide, Mossbauer Effect and its Applications, Tata-Mcgraw Hill, New Delhi,1973
\vskip 0.1cm
[15] J.G.Dash,D.P.Johnson and V.M.Visscher, Phys. Rev., 168(1968)1087

\newpage
Figure captions
\vskip 1.cm
Figure 1:Temperature dependece of Lamb Mossbauer factor and q Lamb Mossbauer factor for various
values of Debye temperature. The curves 'a' and 'b' have been obtained using equation (8) and
curves 'c' and 'd' have been obtained using equation (7). 

\vskip 1.cm
Figure 2:Temperature dependence of Lamb Mossbauer factor and anharmonic Lamb Mossbauer factor for
various values of Debye temperature. The curves 'a' and 'b' have been using equation(8) and curves
'c' and 'd' have been obtained using equation (9).

\newpage
\begin{figure}
\begin{center}
\includegraphics[angle=270,width=16.cm]{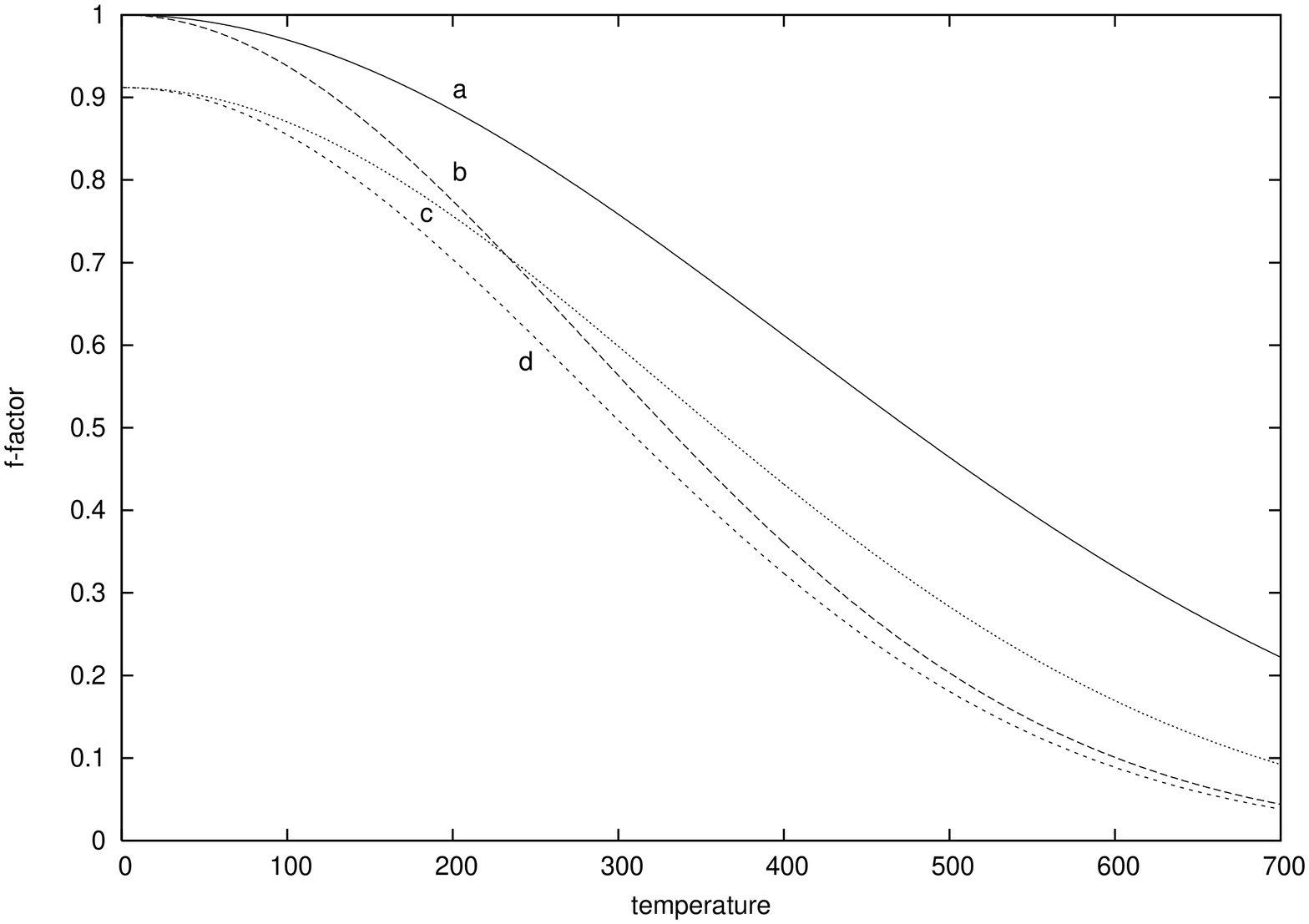}
\caption{}
\end{center}
\end{figure}
\newpage
\begin{figure}
\begin{center}
\includegraphics[angle=270,width=16.0cm]{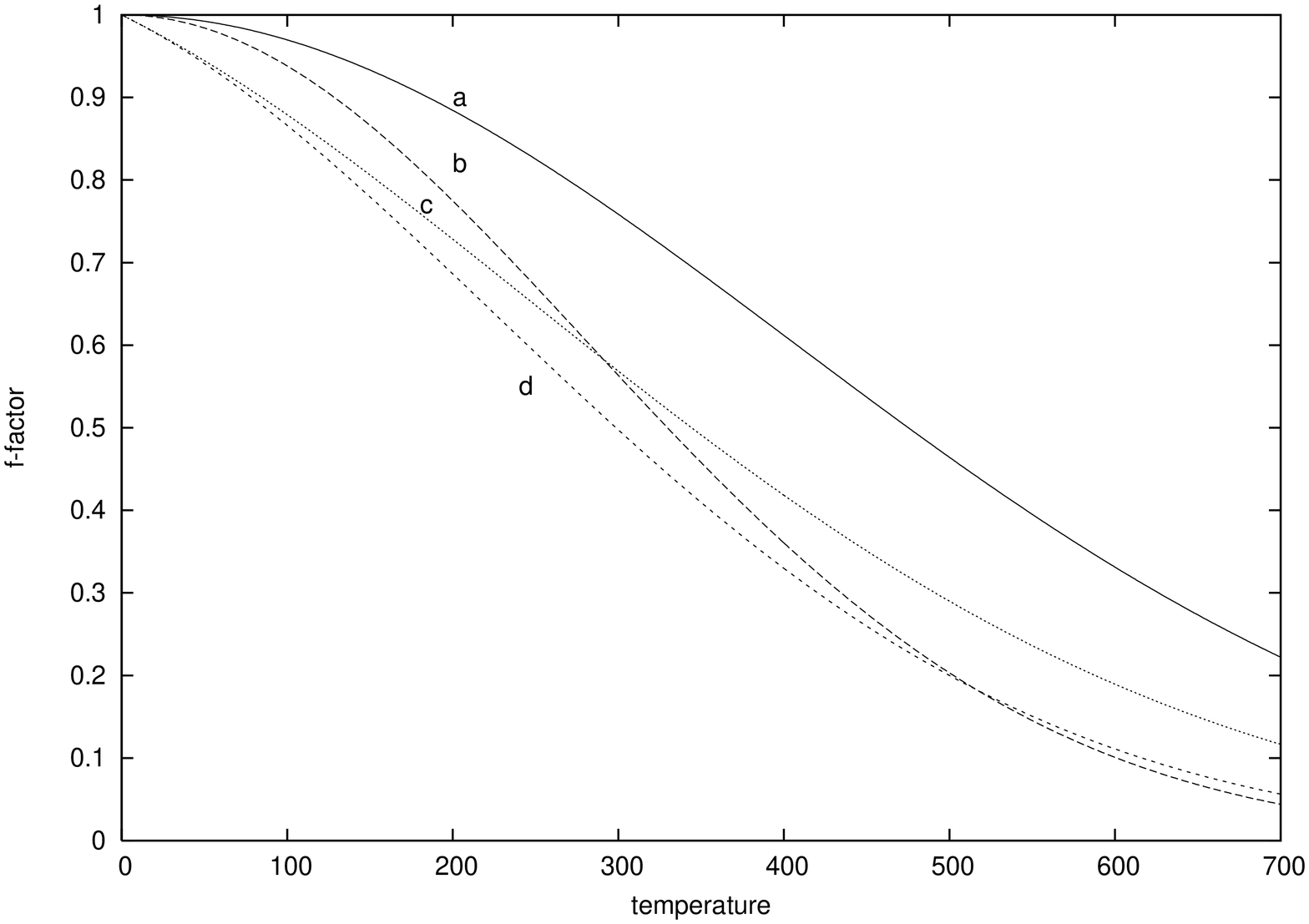}
\caption{}
\end{center}
\end{figure}
\end{document}